\documentclass[aps,prl,twocolumn,superscriptaddress,longbibliography,amsmath,amssymb]{revtex4-2}
\usepackage{amsmath,amssymb,amsfonts}
\usepackage{float}
\usepackage{graphicx} 
\usepackage{dcolumn} 
\usepackage{bm} 
\usepackage{xcolor}
\usepackage{array}
\usepackage{multirow}
\newcolumntype{P}[1]{>{\centering\arraybackslash}p{#1}}
\usepackage{comment}
\usepackage{hyperref}

\hypersetup{
     colorlinks=true,       
     linkcolor=blue,          
     citecolor=black,        
     filecolor=magenta,      
     urlcolor=black         
}

\begin{document}

\title{Highly Polarizable Semiconductors and Universal Origin of
  Ferroelectricity in Materials with a Litharge-Type Structural Unit}

\author{Ziye Zhu}
\affiliation{Department of Materials Science and Engineering, School
  of Engineering, Westlake University, Hangzhou
  310030, China}
\affiliation{Key Laboratory of 3D Micro/Nano Fabrication and Characterization of
Zhejiang Province, School of Engineering, Westlake University, Hangzhou, 310030, China}

\author{Jiaming Hu}
\affiliation{Department of Materials Science and Engineering, School
  of Engineering, Westlake University, Hangzhou
  310030, China}
\affiliation{Key Laboratory of 3D Micro/Nano Fabrication and Characterization of
Zhejiang Province, School of Engineering, Westlake University, Hangzhou, 310030, China}

\author{Yubo Yuan}
\affiliation{Department of Materials Science and Engineering, School
  of Engineering, Westlake University, Hangzhou
  310030, China}
\affiliation{Key Laboratory of 3D Micro/Nano Fabrication and Characterization of
Zhejiang Province, School of Engineering, Westlake University, Hangzhou, 310030, China}

\author{Hua Wang}
\affiliation{Center for Quantum Matter, School of Physics, Zhejiang
  University, Hangzhou, 310058, China}

\author{Xiao Lin}
\affiliation{Key Laboratory for Quantum Materials of
  Zhejiang Province, Department of Physics, School of Science and
  Research Center for Industries of the Future,
  Westlake University, Hangzhou 310030, China}

\author{Wenbin Li}
\email{liwenbin@westlake.edu.cn}
\affiliation{Department of Materials Science and Engineering, School
  of Engineering, Westlake University, Hangzhou
  310030, China}
\affiliation{Key Laboratory of 3D Micro/Nano Fabrication and Characterization of
Zhejiang Province, School of Engineering, Westlake University, Hangzhou, 310030, China}

\begin{abstract}
 We discover that a large family of [Pb$_2$F$_2$]- and
 [Bi$_2$O$_2$]-based mixed-anion materials with a litharge-type
 structural unit are highly polarizable layered semiconductors on the
 edge of ferroelectricity. First-principles calculations demonstrate
 that in this family of materials, compounds as diverse as PbFBr,
 BiOCl, BiCuOSe, Bi$_2$OS$_2$, and Bi$_5$O$_4$S$_3$Cl exhibit static
 dielectric constants an order of magnitude higher than typical
 semiconductors. Additionally, they undergo a ferroelectric transition
 when subjected to a few percent of tensile strain. The ferroelectric
 transitions of these materials are found to have a universal origin
 in the strong cross-bandgap hybridization of the cation $p$ orbitals,
 enabled by the cation 6s$^2$ lone-pair electrons and the
 litharge-type structure of the [Pb$_2$F$_2$] and [Bi$_2$O$_2$]
 layers, as demonstrated by the strain-induced ferroelectric
 transition in the archetypal litharge $\alpha$-PbO. These results
 establish materials with a litharge-type structural unit as a large
 and versatile family of highly polarizable layered semiconductors in
 proximity to ferroelectricity, offering vast opportunities for
 multifunctional materials design.
\end{abstract}

\maketitle

The dielectric response plays a ubiquitous and central role in the
electronic and optical properties of
materials~\cite{Jaramillo2019,Ashcroft1976}. In semiconductors, the
dielectric functions at above-bandgap optical frequencies determines the scale of
linear optical absorption, which is critical for photovoltaic and optical
sensing applications. In the low-frequency regime, the
dielectric polarizability of semiconductors is intricately related to
the screening of bound and mobile charges, affecting properties such
as the dopant ionization energy, insulator-metal transitions, and
carrier lifetime. Indeed, highly polarizable semiconductors that have
a large static (low-frequency) dielectric constant tend
to exhibit longer minority carrier lifetime and enhanced carrier
mobility~\cite{Zhu2016,Zhu2022,He2018}. The possibility of controlling
dielectric response and polar instabilities in highly polarizable
semiconductors through external stimuli such as electric field and
strain may also enable new electronic and photonic
technologies~\cite{Jaramillo2019,Li2021,Wu2023}.

However, semiconductor materials that simultaneously possess a high
dielectric polarizability and excellent electronic properties are
rare. In conventional semiconductors such as Si and GaAs, the static
dielectric constant $\epsilon_0$ typically has a value on the order of
10, where the contribution mainly comes from the polarization of the
electrons~\cite{Madelung2004}. The electronic polarizability
corresponds to the so-called ``high-frequency'' dielectric constant
$\epsilon_{\infty}$, and the difference between $\epsilon_0$ and
$\epsilon_{\infty}$ quantifies the polarizability of the
lattice~\cite{Gonze1997}. Materials with a large static dielectric
constant ($\epsilon_0 > 20$) often fall into the category of large-gap
($>$3~eV) oxide insulators with high lattice polarizability, such as
BaTiO$_3$ and SrTiO$_3$. In recent years, a few highly polarizable
semiconductors with a bandgap in the infrared or visible range were
found in the complex mixed-cation chalcogenide families, including
metal pnictide sulfosalts and the Ba-Zr-S system in the perovskite or
Ruddlesden-Popper structures~\cite{He2018, Filippone2020}. Yet there
was few report of highly polarizable semiconductors that have an
intrinsically layered structure and excellent electrical/optical
characteristics, which can have a significant impact on the current
intense effort of designing next-generation electronic and
optoelectronic devices base on layered and two-dimensional (2D)
materials.

\begin{figure*}[t]
	\centering
	\includegraphics[width=1.0\textwidth]{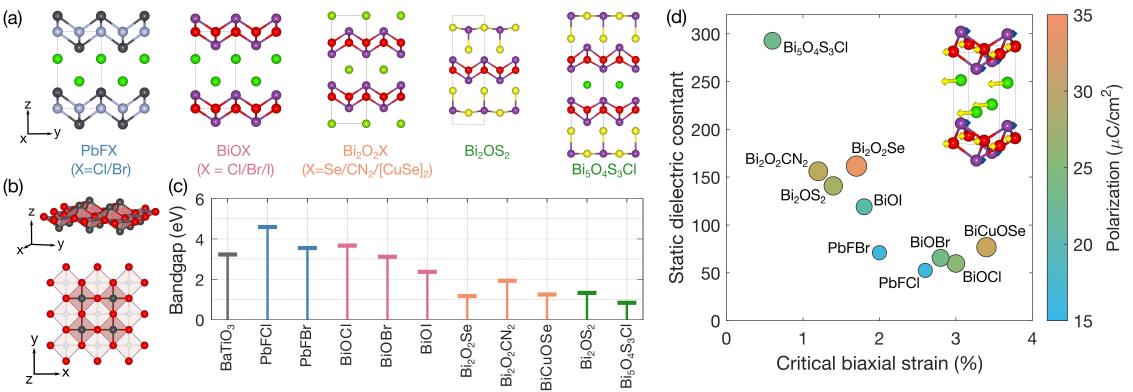}
	\caption{[Pb$_2$F$_2$]- and [Bi$_2$O$_2$]-based highly
          polarizable layered semiconductors and their ferroelectric
          instability. (\textbf{a}) Ball-stick representation of the
          atomistic structures of PbFX (X=Cl/Br), BiOX (X=Cl/Br/I),
          Bi$_2$O$_2$X (X=Se/CN$_2$/[CuSe]$_2$), Bi$_2$OS$_2$, and
          Bi$_5$O$_4$S$_3$Cl. In PbFX, the black, grey and green
          spheres represent Pb, F, and X atoms, respectively. In other
          Bi-based compounds, the Bi and O atoms are represented by
          purple and red spheres, respectively. (\textbf{b}) Side and
          top views of a Pb$_2$O$_2$ layer in $\alpha$-PbO, which has
          the same two-dimensional (2D) litharge-like structure as
          Pb$_2$F$_2$ and Bi$_2$O$_2$ layers. The Pb and O atoms are
          represented by grey and red spheres.  (\textbf{c}) The
          electronic bandgap of the materials calculated by density
          functional theory (DFT) with hybrid functional in the
          Heyd-Scuseria-Enzerhof (HSE) form. The bandgap of BaTiO$_3$
          is also shown for comparison. (\textbf{d}) Calculated static
          dielectric constant of the compounds and their critical
          in-plane biaxial strain for inducing ferroelectric
          instability. The size and color of the markers correspond to
          the electric polarization when the applied in-plane biaxial
          strain is 2\% beyond the critical strain. The inset
          illustrates the directions of atomic displacements
          corresponding to the strain-induced ferroelectric transition
          in BiOCl.}
\label{fig:fig1}
\end{figure*}

In this work, we report the computational discovery of a large family
of [Pb$_2$F$_2$]- and [Bi$_2$O$_2$]-based mixed-anion materials that
are highly polarizable layered semiconductors in proximity to a
ferroelectric transition. This family of materials is formed by
combining [Pb$_2$F$_2$]$^{2+}$ or [Bi$_2$O$_2$]$^{2+}$ layers with
counter-anion layers, leading to a diverse range of compounds
including but not limited to PbFX (X=Cl/Br), BiOX (X=Cl/Br/I),
Bi$_2$O$_2$X (X=Se/CN$_2$/[CuSe]$_2$), Bi$_2$OS$_2$, and
Bi$_5$O$_4$S$_3$Cl. Many of these materials have already been
experimentally synthesized~\cite{Liang2021, Bannister1935, Boller1973,
  Corkett2019, Zhao2014, Phelan2013, Ruan2019, Zou2023, Yang2023}, yet
their unique dielectric peroperties have been overlooked. We found
that all the [Pb$_2$F$_2$]- and [Bi$_2$O$_2$]-based materials exhibit
exceptionally high dielectric polarizability ($\epsilon_0>50$), small
electron effective mass, and highly chemically tunable bandgaps in the
range of \mbox{0--5~eV}, making them attractive for future
electronic and optoelectronic applications.

Importantly, we find that all the [Pb$_2$F$_2$]- and
[Bi$_2$O$_2$]-based layered materials are in proximity to
ferroelectricity, and a ferroelectric transition can occur when an
experimentally accessible tensile strain of 0--3\% is imposed on
them. The ferroelectric transitions are found to have a universal
origin in the strong cross-bandgap hybridization of the cation $p$
orbital, enabled by the cation $6s^2$ lone-pair electrons and the
unique litharge-type structural units in the layered materials. Our
work establishes materials with a litharge-type structural unit and
lone-pair electrons as a large and versatile family of highly
polarizable semiconductors in proximity to a ferroelectric transition,
providing new momentum to the study of beyond-perovskite ferroelectric
systems~\cite{Cheema2020, Cheema2022, Yasuda2024, Bian2024, Cao2024}
that have received tremendous attention recently.

\textsf{\textit{Crystal structure}}. The common structural feature of
the family of materials is the presence of 2D distorted square-net
units of [Bi$_2$O$_2$]$^{2+}$ or [Pb$_2$F$_2$]$^{2+}$ layers, whose
structure bears strong similarity to that of the charge-neutral
[Pb$_2$O$_2$] layers in litharge $\alpha$-PbO, as illustrated in
Figure~\ref{fig:fig1}a and b. Specifically, in a [Bi$_2$O$_2$]$^{2+}$ layer, a sheet of
oxygen arranged in 2D square lattice is sandwiched between two sheets
of bismuth, forming BiO$_4$ square pyramids with the apexes
alternatively pointing in the two layer-normal directions. Similar is
the case for [Pb$_2$F$_2$]$^{2+}$, where Pb and F replace Bi and O,
respectively, and the structure is isoelectronic to
[Bi$_2$O$_2$]$^{2+}$. The nominal charges of Pb and Bi in the layers
are 2+ and 3+, respectively, leaving both Pb$^{2+}$ and Bi$^{3+}$ with
6s$^2$ lone-pair electrons. By stacking [Pb$_2$F$_2$]$^{2+}$ or
[Bi$_2$O$_2$]$^{2+}$ with counter-anion layers such as Cl$^{-}$,
Br$^{-}$, Se$^{2-}$, CN$^{-}$, [CuSe]$_2^{2-}$, [BiS$_2$]$^{-}$, or
[BiS$_3$]$^{3-}$, a wide variety of mixed-anion layered materials can
be formed, which all have a tetragonal lattice with an in-plane
lattice constant in the range of 3.8--4.2~\AA\ (see
Figure~\ref{fig:fig1}a). The bonding within the [Pb$_2$F$_2$]$^{2+}$
or [Bi$_2$O$_2$]$^{2+}$ layers has a strong covalent character,
whereas the bonding between the positively and negatively charged
layers is mostly electrostatic in nature.

\textsf{\textit{Excellent and highly tunable electronic properties.}}
By varying the anion layers, the electronic bandgaps of the
[Pb$_2$F$_2$]- and [Bi$_2$O$_2$]-based materials are highly tunable in
the range of 0--5~eV, as shown in Figure~\ref{fig:fig1}c. For
instance, in Bi$_5$O$_4$S$_3$Cl, which is formed by combining two
[Bi$_2$O$_2$]$^{2+}$ layers with a [BiS$_3$]$^{-}$ layer and a
Cl$^{-}$ layer~\cite{Yang2023}, the bandgap calculated by density
functional theory (DFT) with the Heyd-Scuseria-Ernzerhof (HSE) hybrid
functional~\cite{Heyd2003} has a small value of 0.84~eV. In
comparison, for BiOI and PbFBr, which are formed by stacking
[Bi$_2$O$_2$]$^{2+}$ or [Pb$_2$F$_2$]$^{2+}$ layers with halide anion
layers, the calculated bandgaps have much higher values of 2.37~eV and
3.55~eV, respectively. Importantly, these materials also have highly
dispersive electronic bands. The calculated electron effective mass is
as small as $\sim$0.14$m_0$ in Bi$_2$OS$_2$ and $\sim$0.16$m_0$ in
Bi$_2$O$_2$Se, where $m_0$ is the free-electron mass. In Supplemental
Figure~S1, we present the DFT-calculated band structures of the
[Pb$_2$F$_2$]- and [Bi$_2$O$_2$]-based materials. The methods of
calculation are described in the Supporting Information, and the key
structural and electronic structural information including the lattice
parameters, bandgaps, and electron effective masses, are listed in
Supplemental Table~S1.

\begin{figure*}[t]
	\centering
	\includegraphics[width=1.0\textwidth]{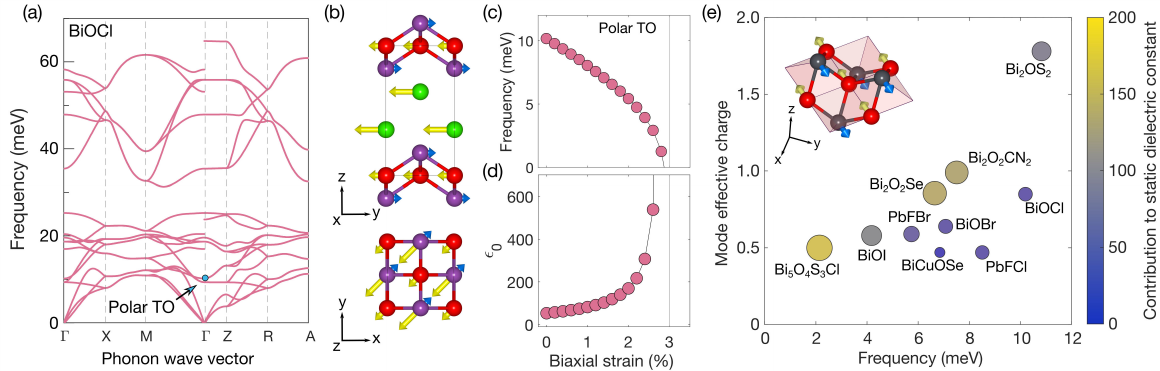}
	\caption{Origin of the high dielectric polarizability of
          [Pb$_2$F$_2$]- and [Bi$_2$O$_2$]-based layered
          semiconductors. (\textbf{a}) Calculated phonon spectrum of
          BiOCl. A low-frequency ($\sim$10~meV) polar transverse
          optical (TO) phonon mode at the $\Gamma$ point is
          indicated. (\textbf{b}) The side and top views of the atomic
          displacement pattern corresponding to the polar TO mode.
          (\textbf{c}) The frequency of the polar TO mode and
          (\textbf{d}) the in-plane static dielectric constant
          ($\epsilon_0$) of BiOCl as a function of in-plane biaxial
          strain. (\textbf{e}) The mode effective charge versus the
          vibrational frequency of the lowest-energy polar TO phonon
          mode in different [Pb$_2$F$_2$]- and [Bi$_2$O$_2$]-based
          materials. The size and color of the markers indicate the
          contribution of the polar TO mode to the static dielectric
          constant. The inset provides a stereo view of the atomic
          displacement pattern corresponding to the TO mode, within
          the [Bi$_2$O$_2$] or [Pb$_2$F$_2$] layers.}
\label{fig:fig2}
\end{figure*}

\textsf{\textit{High polarizability and proximity to ferroelectric
    instability.}} Remarkably, the [Pb$_2$F$_2$]- and
       [Bi$_2$O$_2$]-based materials all have exceptionally high
       dielectric polarizability. As shown in Figure~\ref{fig:fig1}d,
       the in-plane static dielectric constant values $\epsilon_0$,
       calculated for the materials at their experimental lattice
       constants via density functional perturbation theory
       (DFPT)~\cite{Baroni2001}, are all above 50 and can be as high
       as several hundreds. For instance, the $\epsilon_0$ value is
       52.6 in PbFCl, 119.2 in BiOI, and 292.7 in Bi$_5$O$_4$S$_3$Cl,
       which are all an order of magnitude higher than that of silicon
       (12.1) or GaAs (12.8)~\cite{Madelung2004}. By comparing
       $\epsilon_0$ and the corresponding $\epsilon_{\infty}$ values
       listed in Supplemental Table~S1, we conclude that the large
       static dielectric constants of the materials originate from
       high lattice polarizability, which indicates possible
       ferroelectric instability under a slightly changed lattice
       constants.

       Indeed, we find that, by applying a small critical in-plane
       biaxial strain $\varepsilon_{c}$ on the order of 0.5--3.5\%,
       the inversion symmetry of the materials can be broken and a
       ferroelectric phase with non-zero in-plane electric
       polarization emerges. The directions of atomic displacements
       corresponding to the ferroelectric transitions are visualized
       for BiOCl in the inset of Figure~\ref{fig:fig1}d and for other
       materials in Supplemental Figure~S2.  The calculated
       critical biaxial strain $\varepsilon_{c}$ are plotted together
       with $\epsilon_0$ in Figure~\ref{fig:fig1}d, where an overall
       inverse correlation between $\varepsilon_{c}$ and $\epsilon_0$
       can be observed. The strain-dependent potential energy
       landscape of BiOCl with respect to the ferroelectric atomic
       displacements exhibits a classical double-well structure beyond
       $\varepsilon_{c}$, as shown in Supplemental Figure~S3,
       which is consistent with a second-order phase transition. The
       DFT-calculated electric polarizations of the ferroelectric
       phases of the family of materials at 2\% of strain beyond
       $\varepsilon_{c}$ are found to be between
       15--35~$\mu\textrm{C/cm}^2$, which are listed in
       Supplemental Table~S1 and visualized in
       Figure~\ref{fig:fig1}d. In addition to biaxial strain, the
       ferroelectric transition can also be induced by in-plane
       uniaxial strain, and the associated critical strain values are
       slightly larger than those corresponding to biaxial strain and
       listed in Supplemental Table~S1.

\textsf{\textit{Origin of the high polarizability.}} The large family
of [Pb$_2$F$_2$]- and [Bi$_2$O$_2$]-based highly polarizable
semiconductors with excellent electronic properties, as well as their
closeness to ferroelectric instability, prompt us to probe their
fundamental origin. In Figure~\ref{fig:fig2}a we show the calculated
phonon spectrum of BiOCl, which is taken as a representative member of
the material family with $\epsilon_0 = 59.8$. By inspecting the phonon
modes near the center of the Brillouin zone, we find that an in-plane
polar transverse optical (TO) phonon mode has a low frequency of
$\sim$10~meV. By comparison, the lowest-frequency polar TO mode in
GaAs has a much higher frequency of 33~meV~\cite{Madelung2004}. The
displacement pattern of the low-frequency polar TO mode of BiOCl is
illustrated in Figure~\ref{fig:fig2}b, where the Bi atoms move out of
phase with O and Cl atoms. By applying an in-plane biaxial strain, the
polar TO mode rapidly softens (Figure~\ref{fig:fig2}c), accompanied by
a divergent increase of $\epsilon_0$ as the critical point for
ferroelectric instability is approached
(Figure~\ref{fig:fig2}d). Thus, the large $\epsilon_0$ of BiOCl is
strongly correlated with the presence of a low-frequency polar TO
phonon mode.

\begin{figure*}[!t]
	\centering
	\includegraphics[width=1.0\textwidth]{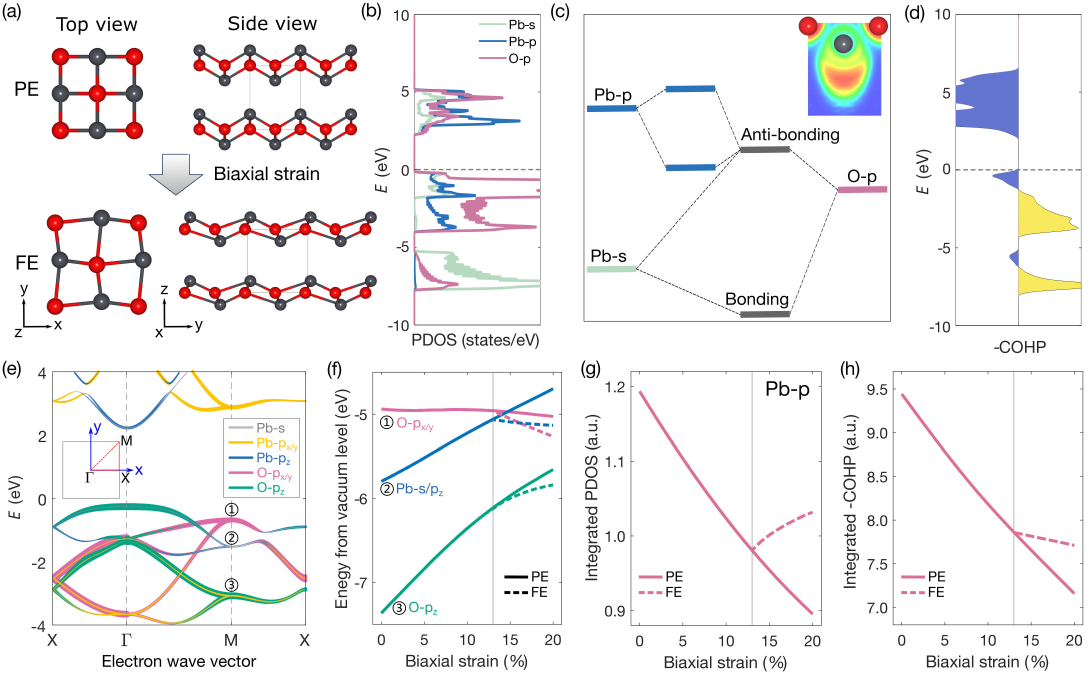}
	\caption{Strain-induced ferroelectricity in archetypal
          litharge $\alpha$-PbO and its fundamental
          origin. (\textbf{a}) The top and side views of the atomistic
          structure of $\alpha$-PbO in the paraelectric (PE) and
          ferroelectric (FE) phase. Pb and O atoms are represented by
          black and red spheres, respectively. (\textbf{b}) Calculated
          projected density of states (PDOS) of monolayer
          $\alpha$-PbO, showing the cross-bandgap hybridization of the
          Pb $p$ orbitals. The energy zero corresponds to the valence
          band maximum (VBM). (\textbf{c}) Molecular orbital diagram
          of $\alpha$-PbO in the revised lone-pair model. The upper
          right corner illustrates the calculated electron
          localization function near a Pb atom of monolayer
          $\alpha$-PbO. Orange and blue represent high and low
          electron localization, respectively. (\textbf{d}) Crystal
          orbital Hamilton population (COHP) of monolayer
          $\alpha$-PbO. The horizontal axis corresponds to
          $-$COHP. (\textbf{e}) DFT-calculated electronic band
          structure of monolayer $\alpha$-PbO. The atomic orbital
          characters of the bands are indicated using different
          colors. (\textbf{f}) Strain-dependent energy levels of the
          electronic states at the M point in the Brillouin
          zone. These states are indicated using the same symbols
          (number in circle) in (\textbf{e}).  The vertical line at
          13~\% of biaxial strain corresponds to the critical strain
          of ferroelectric transition, and the energy-level evolution
          of the PE and FE phases after the critical strain are
          represented using solid and dashed lines,
          respectively. (\textbf{g,h}) The integrated PDOS of Pb $p$
          orbitals and (\textbf{g}) integrated $-$COHP of monolayer
          $\alpha$-PbO (\textbf{h}) as a functional of biaxial
          strain. The upper limit of integration corresponds to the
          VBM. After the critical point, the evolutions of integrated
          PDOS and integrated $-$COHP in both the PE and FE phases are
          shown.}
\label{fig:fig3}
\end{figure*}

More quantitatively, we calculate the contribution of each zone-center
(wavevector $\mathbf{q}=0$) phonon mode to the diagonal components of
the static dielectric tensor $\epsilon_{0,\alpha\alpha}$, which is
proportional to $\Omega^{-1} (Z_{\alpha}/\omega)^2$, where $\alpha$
indicates a Cartesian direction, $\Omega$ is the unit cell volume,
$\omega$ the mode frequency, and $Z_{\alpha}$ is the mode effective
charge vector that reflects the polarity of the zone-center phonon
mode~\cite{Gonze1997}. For a given mode, $Z_{\alpha}$ is calculated as
the sum of the product of the Born effective charge (BEC) tensor
$\mathcal{Z}^*$ and the mass-normalized eigendisplacement for each
atom. For the tetragonal crystals studied here, the in-plane
components of the mode effective charge vector can be written as
$Z_{\alpha, \parallel} = \sum_{\kappa} \mathcal{Z}_{\kappa,
  \parallel}^* \sqrt{M_0/M_{\kappa}} \, e_{\kappa,\alpha} $, where
$\kappa$ is atomic index in the unit cell, $\mathcal{Z}_{\kappa,
  \parallel}$ the in-plane BEC, $M_{\kappa}$ the atomic mass, $M_0$
the mass unit, and $e_{\kappa,\alpha}$ the phonon eigenvector
(detailed discussions can be found in Supplementary Note~S1).

By calculating the $Z$ and $\omega$ for each zone-center phonon mode
of BiOCl, we find that the lowest-frequency polar TO mode contributes
more than 88\% of the total in-plane lattice polarizability of BiOCl
(Supplemental Table~S2). In addition to the small mode frequency
$\omega$ of $\sim$10~meV, the mode effective charge $Z$ of the polar
TO mode has a large value of $\sim$0.85, which is much larger than the
corresponding value of $\sim$0.35 in GaAs. This is because BiOCl has
large BECs, with the in-plane components
$\mathcal{Z}^*_{\parallel}$ equal to $5.34$, $-3.23$, and $-2.11$ for
Bi, O, and Cl, respectively, which are all larger than the nominal
charges of the elements in the compound. For comparison, the
$\mathcal{Z}^*$ calculated for Ga and As in GaAs are 2.1 and $-2.1$
for Ga and As, respectively. In Figure~\ref{fig:fig2}c, we compare the
$Z$ and $\omega$ of the lowest-frequency polar TO mode, as well as the
contribution of the phonon mode to $\epsilon_0$, for all the
[Pb$_2$F$_2$]- and [Bi$_2$O$_2$]-based materials. The vibrational
displacement patterns of the polar optical phonon modes are shown in
Supplemental Figure~S2, and the BECs in different
materials are listed in Supplemental Table~S3. As shown by
Figure~\ref{fig:fig2}e, all the materials have small $\omega$ and
large $Z$ values, and the phonon mode that has a larger $Z/\omega$
value also a larger contribution to the lattice polarizability. Thus,
it is because the existence of a low-frequency polar optical phonon
mode, as well as large BECs (which gives rise to
large mode effective charge), that the [Pb$_2$F$_2$]- and
[Bi$_2$O$_2$]-based materials have exceptionally high lattice
polarizability.

\textsf{\textit{Universal origin of ferroelectric instability}}. We
next investigate why a low-frequency polar TO mode and large BECs are
present in the [Pb$_2$F$_2$]- and [Bi$_2$O$_2$]-based materials, and
why they are susceptible to strain-induced ferroelectric
instability. Noticing that a common feature in these materials is the
presence of [Pb$_2$F$_2$] or [Bi$_2$O$_2$] layered units that bears
strong structural similarity to the charge-neutral [Pb$_2$O$_2$]
layers in the litharge $\alpha$-PbO, we question whether the
strain-induced ferroelectric instability can occur in $\alpha$-PbO as
well. Intriguingly, we find that $\alpha$-PbO can also undergo
ferroelectric transition upon imposing an in-plane biaxial or uniaxial
strain, in both the bulk and monolayer form.

Figure~\ref{fig:fig3}a shows the top and side views of the paraelectric
(PE) phase and biaxial strain-induced ferroelectric (FE) phases of
$\alpha$-PbO. The unique distorted square-net structure in the PE
phase is evident. The DFT-calculated critical biaxial strains to induce
ferroelectric transition in bulk and monolayer $\alpha$-PbO
are both around 13\% (see Supplemental
  Figure~S4). The atomic displacement pattern of ferroelectric
displacement in $\alpha$-PbO bears a strong resemblance to those observed in the
litharge-type layers of the [Pb$_2$F$_2$]- and [Bi$_2$O$_2$]-based
materials.

The calculated projected density of states (PDOS) of
monolayer $\alpha$-PbO, shown in Figure~\ref{fig:fig3}b, exhibits
significant cross-bandgap hybridization of the Pb $p$ orbitals, This
feature can be understood on the basis of the revised lone-pair
model~\cite{Walsh2011}. Due to the high electronegativity of oxygen
and relativity-induced lowering of the Pb $6s$ orbitals, the
energy level of the O $2p$ orbitals is lower than the Pb $6p$ orbitals
but higher than the Pb $6s$ orbitals, which causes Pb cations to exist
in a nominal $2+$ charge state, keeping the $6s^2$ lone-pair
electrons. However, the Pb $6s^2$ electrons are not chemically inert
but interacting strongly with the O $2p$ electrons. This leads to the
formation of (Pb~$6s$ -- O~$2p$) bonding and antibonding orbitals, and
the electrons fill up to the antibonding level.  By distorting the
crystal structure through the displacement of Pb atoms in the out-of-plane
directions of the Pb-O square net, the hybridization of the initially
unoccupied Pb $6p$ orbitals with the (Pb~$6s$ -- O~$2p$) antibonding
orbitals becomes symmetry allowed, resulting in the stabilization of the
occupied antibonding electronic states, as illustrated in
Figure~\ref{fig:fig3}c. The weakly antibonding character of the electronic
states near the valence band maximum (VBM) is confirmed by the crystal
orbital Hamilton population (COHP) analysis in
Figure~\ref{fig:fig3}d.

Thus, it is the presence of $6s^2$ lone-pair electrons in the cations
and the unique distorted bonding geometry of the litharge-type
structure that the Pb $6p$ orbitals have strong cross-bandgap
hybridization. These same bonding conditions are present within the
[Pb$_2$F$_2$] and [Bi$_2$O$_2$] layers of the [Pb$_2$F$_2$]- and
[Bi$_2$O$_2$]-based materials. Indeed, our DFT calculations confirm
that significant cross-bandgap hybridization of the Pb or Bi $6p$
orbitals also exist in the [Pb$_2$F$_2$]- and [Bi$_2$O$_2$]-based
materials, as shown in Supplemental Figure~S5. The same antibonding
characters of the VBM of the materials are further shown in
Supplemental~Figure~S6.

The strong cross-bandgap hybridization leads to large BECs in these
materials, as listed in Supplemental Table~S3. For example, the
in-plane BECs of Pb and Bi in PbFBr and BiOI are 3.97 and 5.92,
respectively, much higher than their respective nominal charges. This
is because relative atomic distortion within the litharge-type layers
of these materials can lead to a substantial change in the covalency
of the occupied electronic states, as the participation of the
unoccupied cation $p$ orbitals in the chemical bonding is susceptible
to the bonding geometry and environment. According to the soft-mode
theory of Cochran~\cite{Cochran1959}, the softening of polar TO mode
is driven by the competition between long-range Coulomb interaction,
which favors a distorted low-symmetry ferroelectric phase, and
short-range forces, which favors the high-symmetry paraelectric
phase. Large BECs lead to a more significant contribution of the
long-range Coulomb interaction, resulting in reduced restoring force
and thus a low-frequency polar TO mode. By applying tensile strain to
the system, the short-range bonding interaction can be weakened, due
to an overall decrease in inter-atomic electronic hybridization. Thus,
as the amount of tensile strain is increased, the frequency of polar
TO modes in materials with large BECs can be further reduced and
eventually become completely softened, resulting in a ferroelectric
transition~\cite{Cohen1992,Cao2024}.

Another perspective to understand the strain-induced ferroelectric
transition is that the ferroelectric displacements of the Pb or Bi
cations can compensate the strain-induced weakening of chemical bonds
through enhanced participation of the unoccupied cation $p$ orbital in
the stabilization of the antibonding valence states. This is supported
by our calculation of the strain-dependent electronic structure
evolution of monolayer $\alpha$-PbO. Figure~\ref{fig:fig3}e shows the
calculated electronic band structure of monolayer $\alpha$-PbO, where
three electronic state at the M-point are indicated. As the imposed
in-plane biaxial strain is increased, it is observed that the
electronic states with initially substantial admixture of Pb $p$
orbitals rapidly rise in energy (Figure~\ref{fig:fig3}f), as the
stabilization effect of the Pb $p$ orbital is weakened due to
increased interatomic distances. This is reflected in
Figure~\ref{fig:fig3}g and Figure~\ref{fig:fig3}h, where rapid
decreases in the calculated integrated PDOS of Pb $p$ orbitals in the
valence manifold and the integrated $-$COHP values can be seen. Beyond
the critical strain of $\sim$13\%, the lateral distortion of the Pb
atoms in the ferroelectric phase can reverse the trend of the
participation of Pb $p$ orbitals in the bonding
(Figure~\ref{fig:fig3}g), which leads to a slowed weakening in the total
bonding strength (Figure~\ref{fig:fig3}h), as well as energy gain with
respect to the paraelectric phase.

Compared to monolayer $\alpha$-PbO, the mixed-anion, quasi-layered
structure of the [Pb$_2$F$_2$]- and [Bi$_2$O$_2$]-based materials with
additional counter-anion layers in between the [Pb$_2$F$_2$] or
[Bi$_2$O$_2$] layers further reduces the overall short-range forces
that inhibit the softening of the polar TO phonons. As shown in
Figure~\ref{fig:fig2}b and Supplemental Figure~S2, the polar TO
phonons in these materials involve not only the internal displacements
within the [Pb$_2$F$_2$] or [Bi$_2$O$_2$] layers, but also the
relative displacements between the litharge-type layers and the
counter-anion layers. Unlike the strong covalent bonding within the
litharge-type [Pb$_2$F$_2$] or [Bi$_2$O$_2$] layers, the bonding
between the counter-anion layers and the litharge-type layers has a
weak covalent character and is more electrostatic in origin.  In
addition, the smaller bandgaps of the [Bi$_2$O$_2$]-based materials
lead to their stronger cross-bandgap hybridization (Supplemental
Figure~S5), which results in larger BECs (Supplemental Table~S3) and
stronger long-range Coulomb interactions. Consequently, the
[Pb$_2$F$_2$]- and [Bi$_2$O$_2$]-based materials have a reduced critical
tensile strain (between 0.5\% to 3\%) needed for the complete
softening of the polar TO mode and thus the ferroelectric transition.

\textsf{\textit{Conclusions and outlook}}. In conclusion, we discover
that a large family of [Pb$_2$F$_2$]- and [Bi$_2$O$_2$]-based
mixed-anion materials with a litharge-type structural unit are highly
polarizable layered semiconductors with structural diversity, a wide
range of bandgaps (0--5~eV), and excellent electronic properties,
making them attractive for applications in electronics and
optoelectronics based on layered and 2D materials. We further uncover
the fundamental origin of their high lattice polarizability by
revealing the crucial roles of cation $6s^2$ lone-pair electrons and
cross-bandgap electronic hybridization in the presence of a
low-frequency polar TO phonon mode in their lattice dynamics. Our
results indicate that proximity to strain-induced ferroelectricity is
a universal phenomenon in [Pb$_2$F$_2$]- and [Bi$_2$O$_2$]-based
compounds with a litharge-type structural unit, opening a large
chemical space for designing novel ferroelectric and incipient
ferroelectric materials, whose bandgaps can be varied from infrared to
ultraviolet spectrum by choosing proper counter-anion layers.

Importantly, the crystal symmetries and lattice constants of the
[Bi$_2$O$_2$]- and [Pb$_2$F$_2$]-based materials have close match with
each other and with the perovskite family of materials such as
SrTiO$_3$ and LaAlO$_3$. Thus, a large variety of novel
heterostructures can be constructed by stacking [Bi$_2$O$_2$]- and
[Pb$_2$F$_2$]-based materials with each other, or with the perovskite
oxides, offering rich heterostructural design freedom, especially
given the highly chemically tunable bandgaps of the [Bi$_2$O$_2$]- and
[Pb$_2$F$_2$]-based materials. Due to the close proximity of the
materials to ferroelectric transition, the mismatch strain at the
heterostructural interface could be sufficient to induce
ferroelectricity, while the polar discontinuity at the
heterostructural interface may causes carrier doping~\cite{Zhu2023},
which could lead to the coveted polar metallic
phase~\cite{Anderson1965, Bhowal2023} in such
heterostructures. Notably, in degenerately doped [Bi$_2$O$_2$]-based
compounds, superconductivity has been experimentally
reported~\cite{Phelan2013,Ruan2019,Zou2023,Yang2023}. Thus, it is also
possible that the long-sought coexistence of ferroelectricity and
superconductivity could be realized in this large family of materials
and their heterostructures. In summary, the [Bi$_2$O$_2$]- and
[Pb$_2$F$_2$]-based materials with a litharge-type structural unit
offer an exciting playground for designing high performance and highly
polarizable layered semiconductors, as well as novel ferroelectric
heterostructures, with vast potential to uncover exotic solid state
phenomena for emerging technologies.

\textsf{\textit{Acknowledgments}}. The work of Z.Z., J.H., Y.Y., and
W.L. is supported by the National Natural Science Foundation of China
(NSFC) under Grant No. 62374136. W.L. also acknowledges the support by
Research Center for Industries of the Future at Westlake University
under Award No. WU2022C041. X.L. acknowledges the support by
``Pioneer'' and ``Leading Goose'' R$\&$D Program of Zhejiang under
Grant 2024SDXHDX0007, Zhejiang Provincial Natural Science Foundation
of China for Distinguished Young Scholars under Grant No. LR23A040001,
and the NSFC under Grant No. 12474131. H.W. acknowledges the support
from the NSFC under Grant Nos. 12304049 and 12474240. The authors
thank Prof. S.H. Wei for inspiring discussions and the HPC center of
Westlake University for computational support. Z. Zhu and J. Hu
contributed equally.

%

\end{document}